\documentclass[aps,pre,twocolumn,showpacs,superscriptaddress]{revtex4}
\usepackage{epsfig}
\usepackage{times}
\usepackage{amsmath}
\begin{document}

\title{From pairwise to group interactions in games of cyclic dominance}

\author{Attila Szolnoki}
\affiliation{Institute of Technical Physics and Materials Science, Research Centre for Natural Sciences, Hungarian Academy of Sciences, P.O. Box 49, H-1525 Budapest, Hungary}

\author{Jeromos Vukov}
\affiliation{Institute of Technical Physics and Materials Science, Research Centre for Natural Sciences, Hungarian Academy of Sciences, P.O. Box 49, H-1525 Budapest, Hungary}

\author{Matja{\v z} Perc}
\affiliation{Faculty of Natural Sciences and Mathematics, University of Maribor, Koro{\v s}ka cesta 160, SI-2000 Maribor, Slovenia}

\begin{abstract}
We study the rock-paper-scissors game in structured populations, where the invasion rates determine individual payoffs that govern the process of strategy change. The traditional version of the game is recovered if the payoffs for each potential invasion stem from a single pairwise interaction. However, the transformation of invasion rates to payoffs also allows the usage of larger interaction ranges. In addition to the traditional pairwise interaction, we therefore consider simultaneous interactions with all nearest neighbors, as well as with all nearest and next-nearest neighbors, thus effectively going from single pair to group interactions in games of cyclic dominance. We show that differences in the interaction range affect not only the stationary fractions of strategies, but also their relations of dominance. The transition from pairwise to group interactions can thus decelerate and even revert the direction of the invasion between the competing strategies. Like in evolutionary social dilemmas, in games of cyclic dominance too the indirect multipoint interactions that are due to group interactions hence play a pivotal role. Our results indicate that, in addition to the invasion rates, the interaction range is at least as important for the maintenance of biodiversity among cyclically competing strategies.
\end{abstract}

\pacs{02.50.+Le, 07.05.Tp, 89.75.Fb}

\maketitle

\section{Introduction}
Rock is wrapped by paper, paper is cut by scissors, and scissors are broken by a rock. This is the simple blueprint of the rock-paper-scissors game, where the three competing strategies form a closed loop of dominance. The game is popular among children and adults to decide on trivial disputes that have no obvious winner, but it is also the basis for the explanation of fascinating evolutionary processes that describe the essence of predator-prey interactions and evolutionary games \cite{maynard_82, hofbauer_98, mestertong_01, nowak_06}. Foremost, games of cyclic dominance play a prominent role in explaining the intriguing diversity in nature \cite{frachebourg_prl96, mobilia_pre06, reichenbach_n07, kerr_n02, reichenbach_prl07, mobilia_jtb10, reichenbach_prl08}, but they are also able to provide insights into Darwinian selection \cite{maynard_n73} as well as structural complexity \cite{watt_je47} and pre-biotic evolution \cite{rasmussen_s04}.

Cyclic interactions may also arise spontaneously in evolutionary games where the number of competing strategies is three or more. For example, cyclic dominance has been observed in public goods games with volunteering \cite{hauert_s02, semmann_n03}, peer punishment \cite{hauert_s02, helbing_ploscb10, amor_pre11}, pool punishment \cite{szolnoki_pre11, sigmund_n10} and reward \cite{szolnoki_epl10, sigmund_pnas01}, but also in pairwise social dilemmas with coevolution \cite{szolnoki_epl09, szolnoki_pre10b} or with jokers \cite{requejo_pre12b}. The ample attention to the theoretical aspects of cyclical interactions is fueled by actual observations of such interactions in nature. Prominent examples include the mating strategy of side-blotched lizards \cite{sinervo_n96}, overgrowth of marine sessile organisms \cite{burrows_mep98}, genetic regulation in the repressilator \cite{elowitz_n00}, and competition in microbial populations \cite{durrett_jtb97, kirkup_n04, neumann_gf_bs10, nahum_pnas11}.

The key to the sustenance of biodiversity, in addition to the inherent closed loop of dominance that governs the evolution of such systems, is often spatial structure \cite{szabo_pr07, perc_jrsi13, jiang_pre11}. In experiments with \textit{Escherichia coli}, for example, it has been shown that arranging the bacteria on a Petri dish is crucial for keeping all three competing strains alive \cite{kerr_n02, kerr_n06}. Accordingly, simulations of spatial rock-paper-scissors and related games of cyclic dominance have a long and fruitful history \cite{szabo_pre99, frean_prsb01, reichenbach_pre06, szabo_jtb07, peltomaki_pre08, peltomaki_pre08b, berr_prl09, he_q_pre10, wang_wx_pre10b, ni_x_c10, wang_wx_pre11, mathiesen_prl11, avelino_pre12, jiang_ll_pla12, roman_jsm12, avelino_pre12b, juul_pre12, roman_pre13, vukov_pre13, laird_oikos14}, which is firmly rooted in methods of statistical physics. In general, if the mobility in the population is sufficiently small \cite{reichenbach_n07}, the spatial rock-paper-scissors game leads to the stable coexistence of all three competing strategies, whereby the coexistence is maintained by the spontaneous formation of complex spatial patterns. Most intriguing recent examples of this phenomenon include the observation of labyrinthine clustering \cite{juul_pre13} and interfaces with internal structure \cite{avelino_arx13}.

Here we wish to extend the scope of the spatial rock-paper-scissors game by abandoning the common assumption of single pairwise interactions. Traditionally, when two players are randomly selected from the population, one is able to invade the other based on the governing food web and the invasion rates \cite{perc_pre07b, szabo_pre08}. In reality, however, the fitness of each individual player depends not only on one nearest neighbor, but may be influenced by all nearest neighbors and beyond. Going from pairwise to simultaneous interactions in a group is known to vitally affect the outcome of evolutionary games \cite{perc_jrsi13}, but this transition has been neglected in games of cyclic dominance. To overcome this, we consider that the invasion rates determine individual payoffs, which in turn govern the process of strategy change. The traditional version of the game is recovered if the payoffs for each potential invasion stem from a single pairwise interaction, but the game differs if we consider interactions with nearest and next-nearest neighbors for the accumulation of payoffs. As we will show, the differences in the interaction range can decelerate as well as revert the invasion between the competing strategies, thus qualitatively changing the evolutionary outcome that would be expected based on the pairwise consideration of the governing food web. Indirect multipoint interactions that are due to simultaneous interactions within a group thus play a pivotal role in games of cyclic dominance, similarly as reported before for evolutionary social dilemmas \cite{szolnoki_pre09c}.

The organization of this paper is as follows. We present the definition of the model and the elementary results that hold under well-mixed conditions in Section~II. Main results are presented in Section~III. We conclude with the summary of the results and a discussion of their implications in Section~IV.

\section{Model definition}
\begin{figure}
\centerline{\epsfig{file=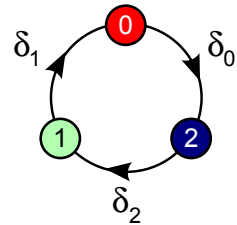,width=3cm}}
\caption{(Color online) Schematic presentation of the food web that describes the closed loop of dominance between the three competing strategies. Just like rock is wrapped by paper, paper is cut by scissors, and scissors are broken by a rock, here strategies $0$ (medium gray), $1$ (light gray) and $2$ (dark gray) form a closed loop of dominance. If the game is governed by individual pairwise interactions, the probabilities of invasion are determined by $\delta_0$, $\delta_1$ and $\delta_2$, respectively. If the game is governed by group interactions, however, these probabilities are transformed into payoffs that may dictate altogether different evolutionary outcomes.}
\label{foodweb}
\end{figure}

We consider the classic rock-paper-scissors game, where the three strategies cyclically dominate each other, as depicted schematically in Fig.~\ref{foodweb}. For convenience, we refer to the strategies as $0$, $1$ and $2$, where strategy $0$ invades strategy $2$ with probability $\delta_0$, strategy $1$ invades strategy $0$ with probability $\delta_1$, and strategy $2$ invades strategy $1$ with probability $\delta_2$.

The stationary state can be described by the fractions of strategies where $\rho_i$ denotes the fraction of strategy $i \in (0,1,2)$ in the population. In a well-mixed system, the time dependence of $\rho_i$ is described by the following system of differential equations
\begin{equation}
\dot{\rho}_i = \delta_{i} \rho_i \rho_{i+2} - \delta_{i+1} \rho_i \rho_{i+1}\,\,,
\end{equation}
where $i$ runs from $0$ to $2$ in a cyclic manner. Due to the defined dynamics, the sum of all $\rho_i$ is conserved and is always equal to $1$. The stationary values of $\rho_i$ are easily obtained according to
\begin{equation}
\rho_i = \frac{\delta_{i+2}}{\sum_i \delta_i}\,\,,
\label{statfreq}
\end{equation}
where $i$ again runs from $0$ to $2$ cyclically. This result indicates that the fractions of strategies are determined unambiguously by the three invasion rates. As reviewed in the Introduction, however, it is frequently assumed that the interactions among the competing strategies, which are summarized by the corresponding invasion rates, are more complex and in fact are described by a structured population. It is known that the transition from a well-mixed to a structured population affects the evolutionary outcome of the rock-paper-scissors game in both qualitative and quantitative ways. The question we seek to answer is whether the same holds on structured populations if we go from individual pairwise interactions to simultaneous group interactions.

For the spatial version of the game, we consider that each player is located on the site $x$ of a square lattice with periodic boundary conditions, where the grid contains $L \times L$ sites. The three strategies are initially distributed uniformly at random with no sites left empty. We adopt the strategy notation of three-dimensional unit vectors

\begin{equation*}
{\bf s}_x =0= \begin{pmatrix}1 \\0 \\ 0 \end{pmatrix}, \textrm{or }
{\bf s}_x =1= \begin{pmatrix}0 \\1 \\ 0 \end{pmatrix}, \textrm{or }
{\bf s}_x =2= \begin{pmatrix}0 \\0 \\ 1 \end{pmatrix}.
\end{equation*}

Accordingly, the payoff of player $x$ against the neighbor at site $y$ can be expressed by the following matrix product
\begin{equation}
\Pi_x= {\bf s}^{+}_x {\bf M} \cdot {\bf s}_{y}\,,
\end{equation}
where ${\bf s}^{+}_x$ denotes the transpose of the state vector ${\bf s}_x$. In the present case, the payoff matrix is given by

\begin{equation}
{\bf M}=\begin{pmatrix} \phantom{-}0 & -\delta_1 & \phantom{-}\delta_0 \\
                        \phantom{-}\delta_1 & \phantom{-}0 & -\delta_2  \\
                       -\delta_0 & \phantom{-}\delta_2 & \phantom{-}0
\end{pmatrix}.
\end{equation}

The evolution of strategies proceeds in agreement with a random sequential update, where during a full Monte Carlo step every player receives a chance once on average to invade one randomly selected neighbor. In particular, a randomly selected player $x$ will invade the neighbor $y$ with the rate proportional to $\frac{\Pi_x - \Pi_y}{2}$, where the acquisition of the payoffs $\Pi_x$ and $\Pi_y$ depends on the type of interaction that governs the evolutionary process. We will refer to an interaction as a ``pair'' interaction when the two competing players interact only with each other. As already noted, this returns the traditional version of the spatial rock-paper-scissors game, where the payoffs correspond directly to the invasion rates defined in Fig.~\ref{foodweb}. Moreover we will consider two additional cases, where the interaction range is gradually extended from the mentioned two-player interaction via the von~Neumann neighborhood, entailing four pair interactions with the nearest neighbors, to the Moore neighborhood, entailing eight pair interactions. In the latter case, a player interacts with its four nearest as well as with four next-nearest neighbors simultaneously. This transition from individual pairwise to group interactions for the determination of payoffs is the key aspect of the current work.

In the following Section, for the sake of simplicity, we fix the invasion rate $\delta_2=1$, thus $\delta_0$ and $\delta_1$ remain as the two free parameters. For technical reasons, we also always keep both parameters within the interval $0.2 \le \delta_0, \delta_1 \le 1$ in order to avoid heavy finite size effects reported in \cite{juul_pre13}. By doing so, we achieve that during the Monte Carlo simulations the $L \times L = 400 \times 400$ system size is always sufficiently large to get reliable results that are valid also in the large size limit. Our main conclusions, however, are robust and remain valid in the whole $(\delta_0,\delta_1) \in [0,1]$ interval as well.

\section{Results}

\begin{figure}
\centerline{\epsfig{file=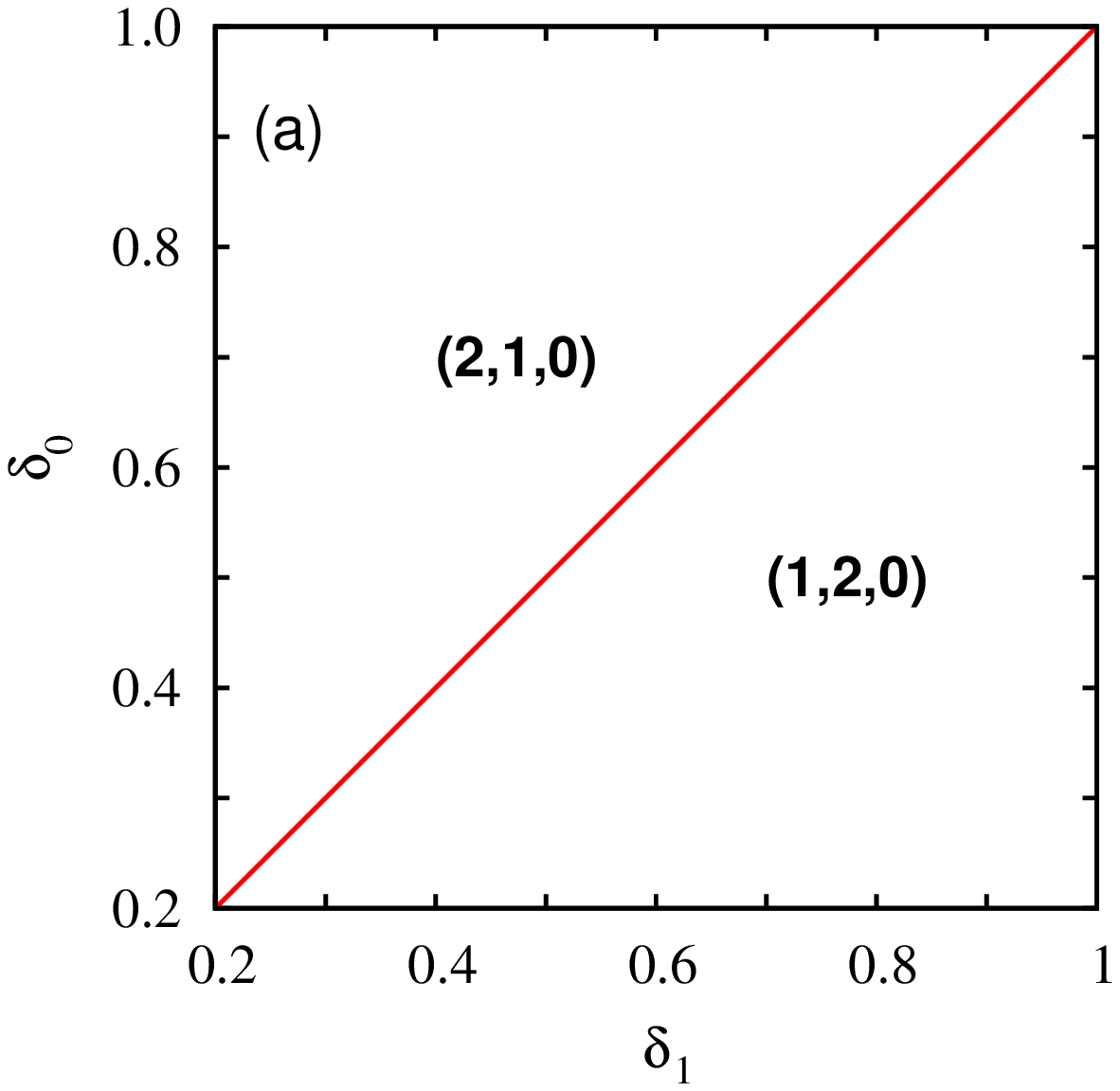,width=4.25cm}\epsfig{file=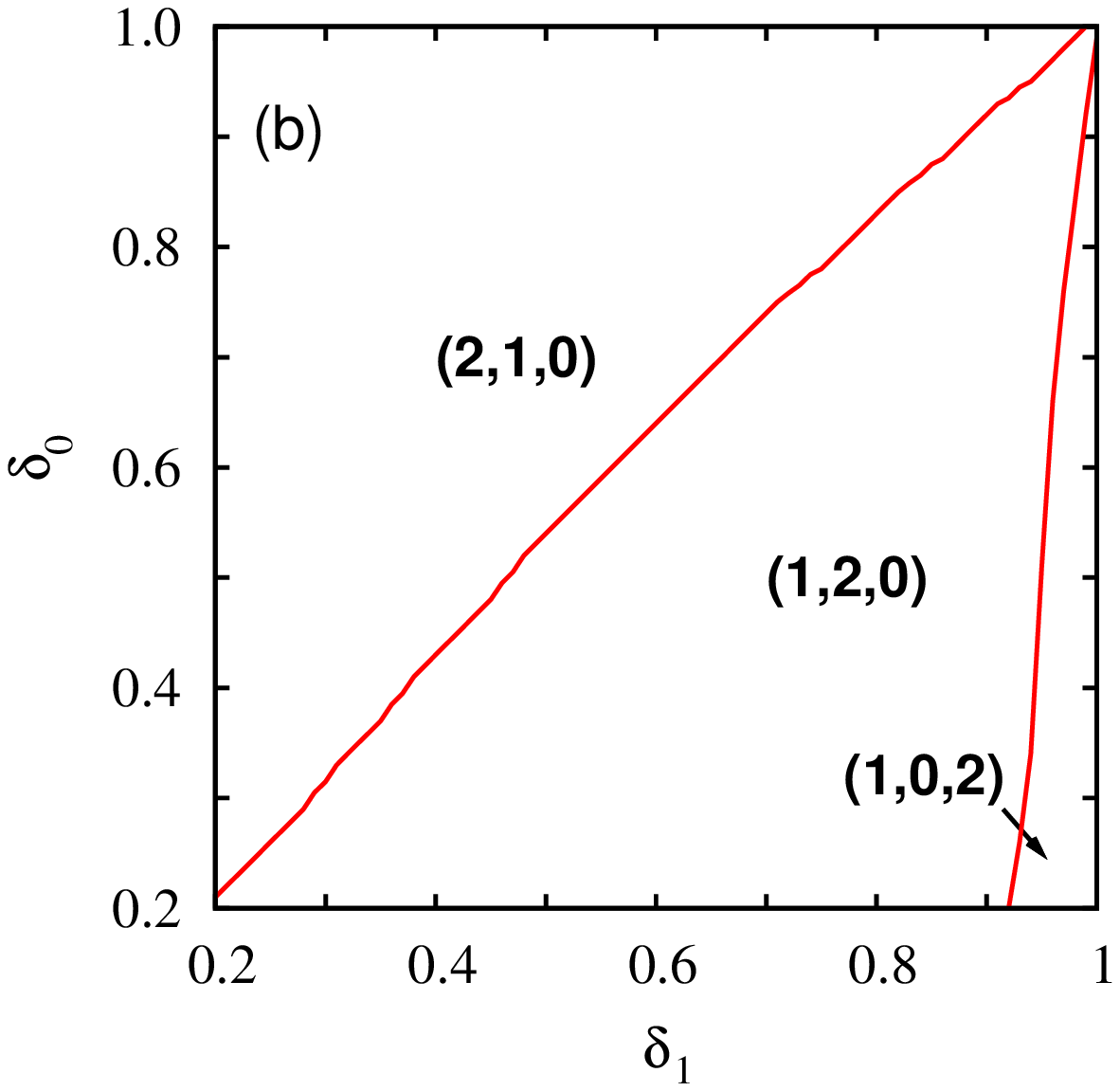,width=4.25cm}}
\centerline{\epsfig{file=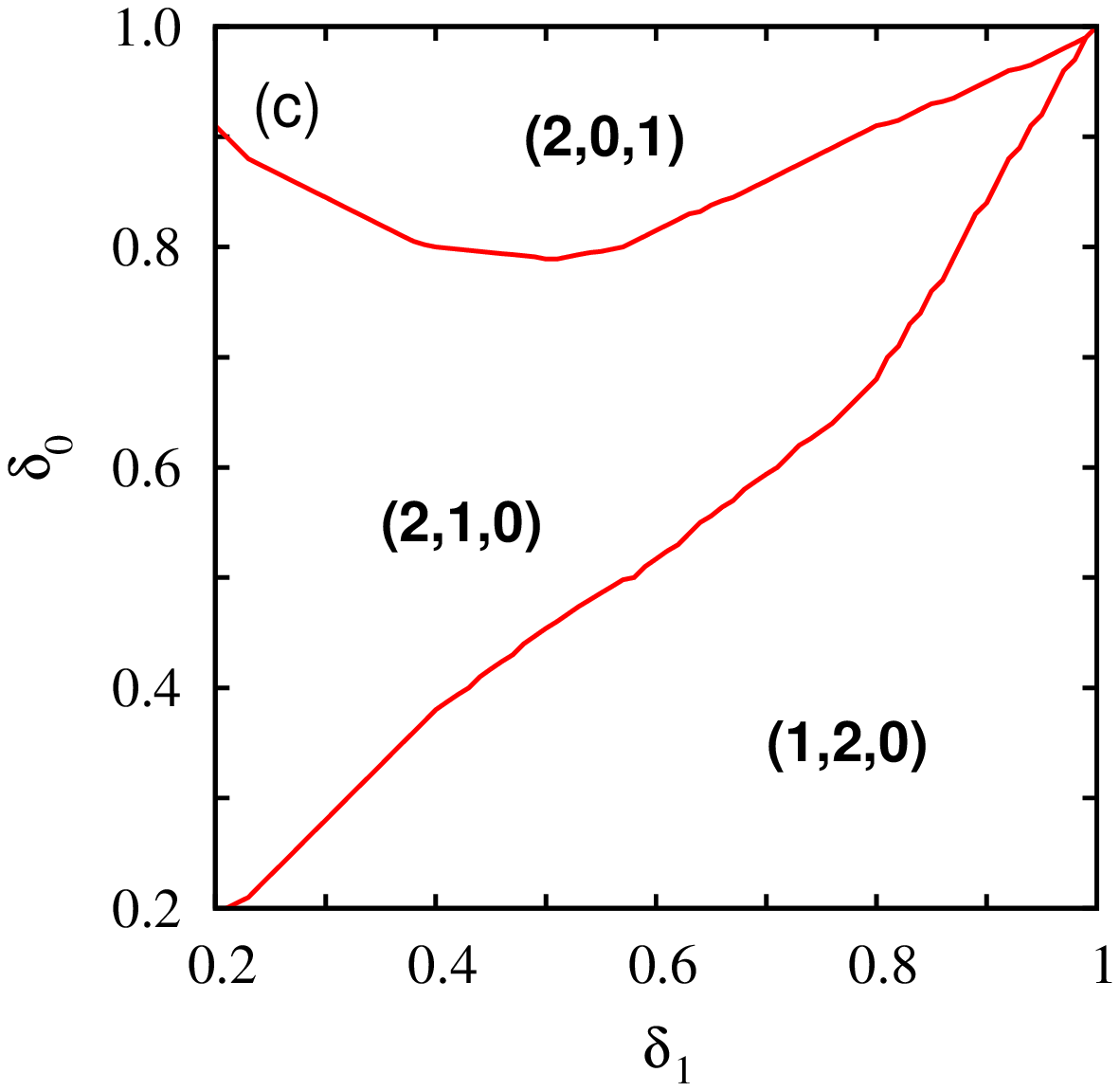,width=4.25cm}\epsfig{file=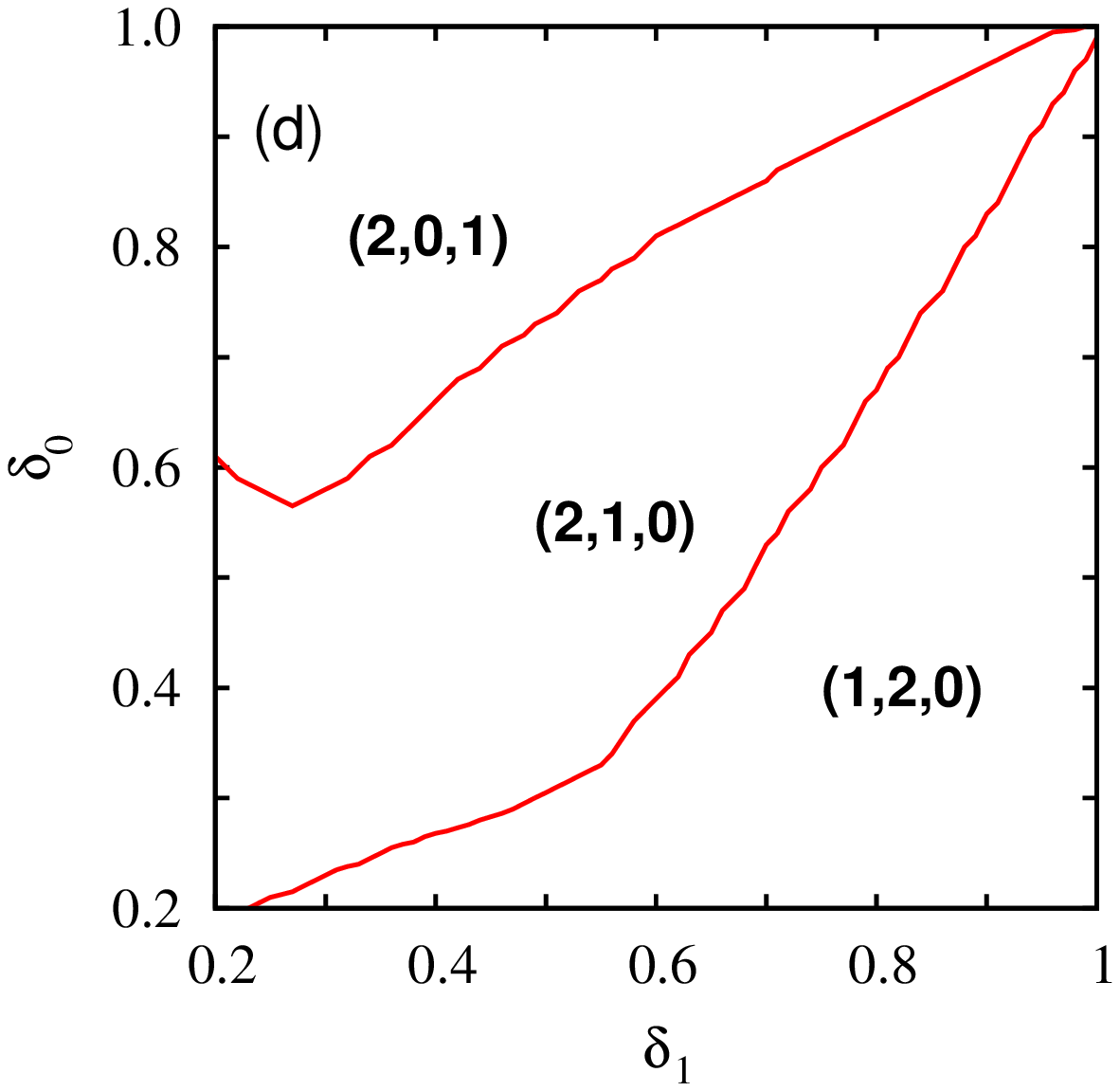,width=4.25cm}}
\caption{(Color-online) Phase diagrams in dependence on the $1 \to 0$ invasion rate $\delta_1$ and the $0 \to 2$ invasion rate $\delta_0$, as obtained for (a) well-mixed interactions, (b) ``pair'' interactions, (c) von~Neumann interactions, and (d) Moore interactions. Phases are defined according to the rank of strategy fractions in the stationary state. The $(2,1,0)$ phase, for example, corresponds to $\rho_2<\rho_1<\rho_0$. The $2 \to 1$ invasion rate is fixed to $\delta_2=1$.}
\label{phase}
\end{figure}

We begin by showing phase diagrams for different interaction ranges in Fig.~\ref{phase}, where the phases are defined according to the rank of strategy fractions in the population. The $(2,1,0)$ phase thus corresponds to a stationary state where $\rho_2<\rho_1<\rho_0$. The $(1,2,0)$ phase, on the other hand, corresponds to $\rho_1<\rho_2<\rho_0$. In panel (a) of Fig.~\ref{phase}, we show the phase diagram obtained in the well-mixed model. Since the invasion rate $\delta_1$ determines the $1 \to 0$ invasion and the invasion rate $\delta_0$ determines the $0 \to 2$ invasion, according to Eq.~\ref{statfreq} the expected result is a right diagonal across the $\delta_1 - \delta_0$ plane that delineates the $(2,1,0) \to (1,2,0)$ phase transition. Replacing well-mixed interactions with ``pair'' interactions on the square lattice leads to expected quantitative and qualitative changes. As depicted in panel (b) of Fig.~\ref{phase}, the $(2,1,0) \to (1,2,0)$ diagonal becomes slightly upward bent, but more importantly, a new $(1,0,2)$ phase emerges for large $\delta_1$ values that is missing under well-mixed conditions.

Significantly more unexpected is the observation that replacing ``pair'' interactions with von~Neumann interactions and further with Moore interactions introduces further quantitative as well as qualitative changes. These changes are no longer due to the transition from well-mixed to structured populations, but are the sole consequence of different interaction ranges. As depicted in panels (c) and (d), a new $(2,0,1)$ phase emerges for large $\delta_0$ values that is absent under well-mixed conditions and ``pair'' interactions, moreover, the $(1,0,2)$ phase that we have observed for ``pair'' interactions vanishes. In addition to these qualitative changes that are due to the transition from pairwise to group interactions, quantitative changes responsible for the shifts in the position of the phase transitions on the $\delta_1 - \delta_0$ plane are also clearly inferable.

\begin{figure}
\centerline{\epsfig{file=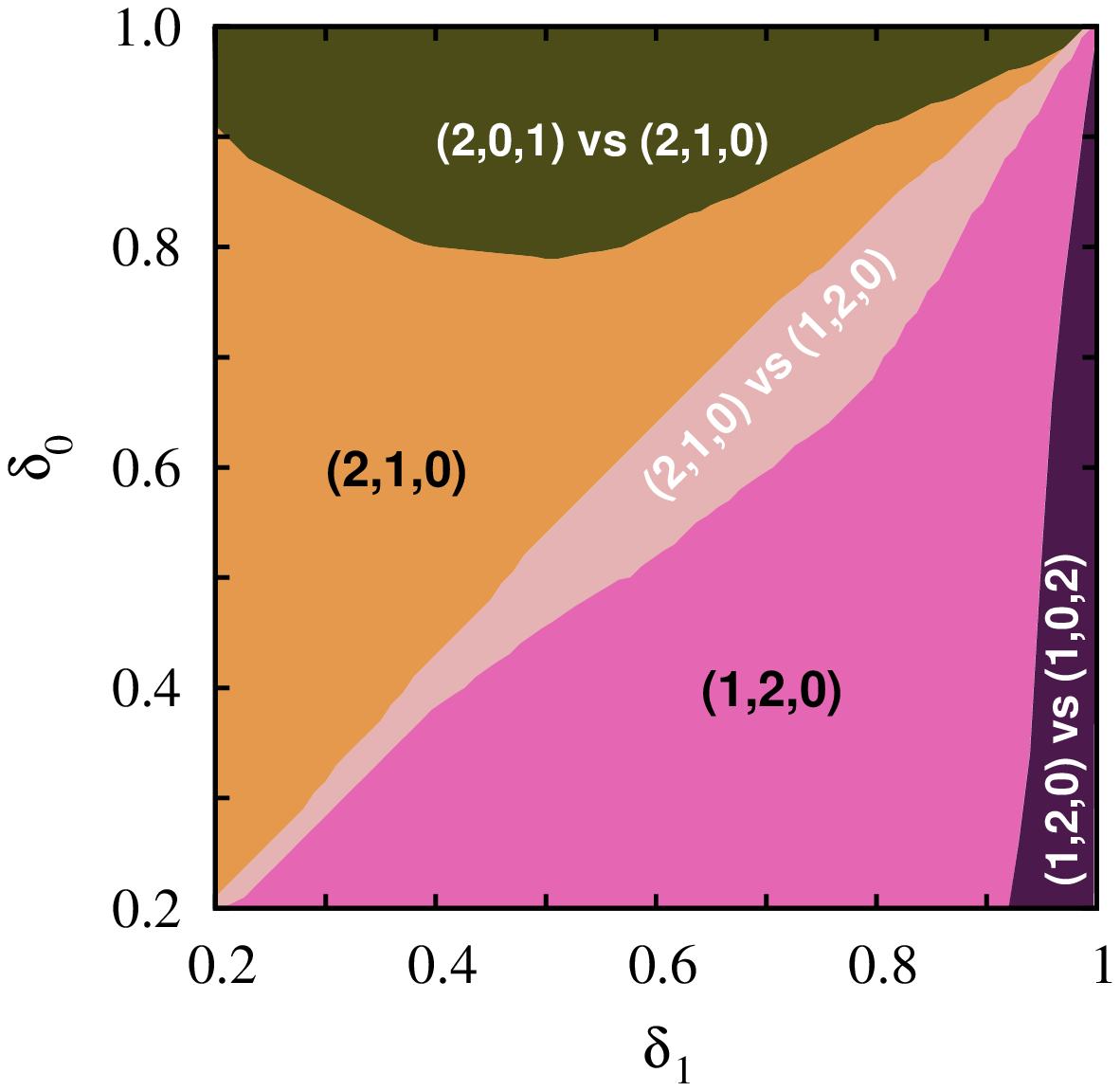,width=7cm}}
\centerline{\epsfig{file=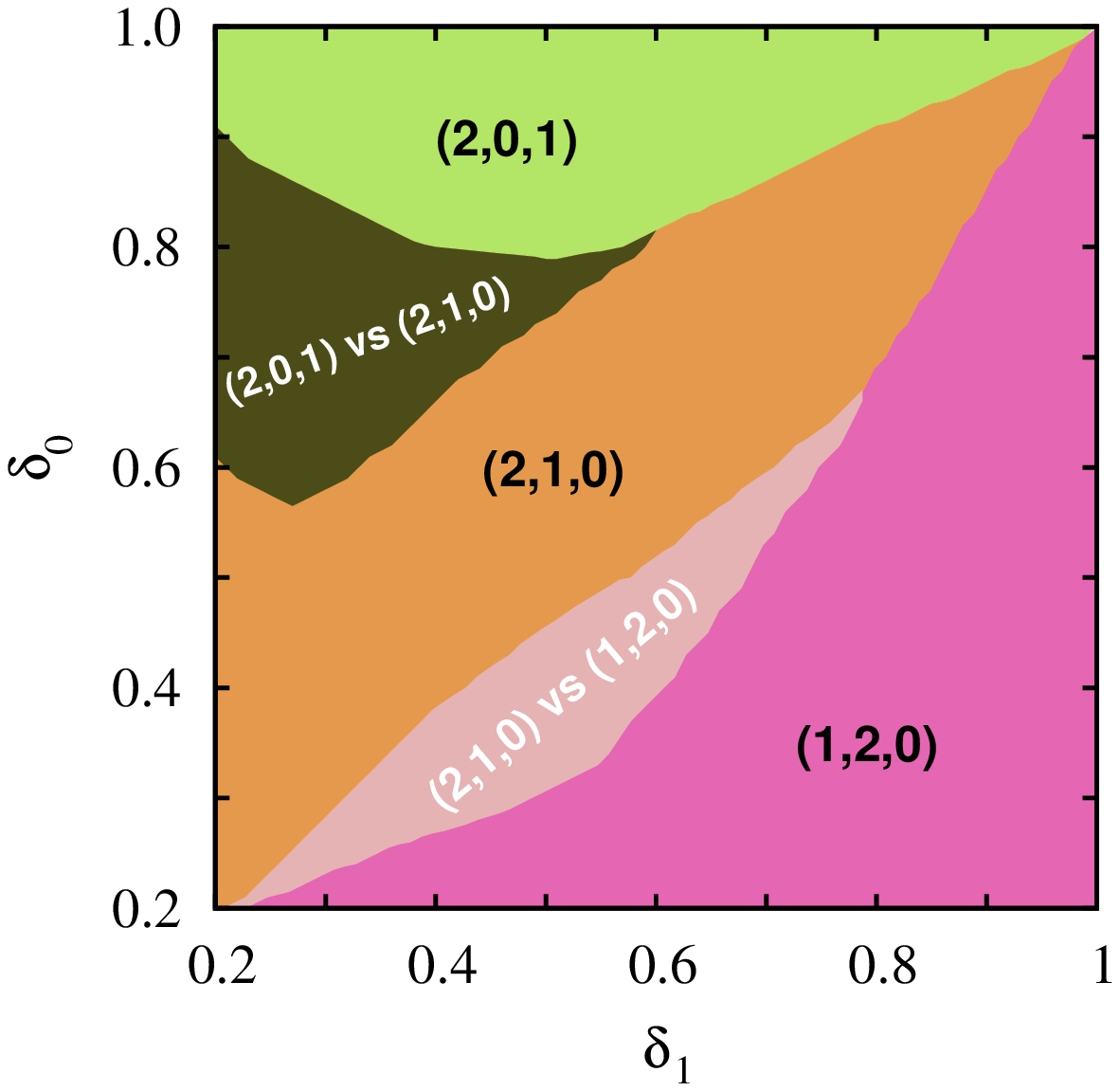,width=7cm}}
\caption{(Color online) Differences between the phase diagrams depicted in Fig.~\ref{phase}. The upper panel depicts the differences between ``pair'' and von~Neumann interactions [panels (b) and (c) in Fig.~\ref{phase}], while the lower panel depicts the differences between von~Neumann and Moore interactions [panels (c) and (d) in Fig.~\ref{phase}]. Regions in the $\delta_1 - \delta_0$ plane where different interaction ranges yield different phases (qualitative differences) are labeled white, while regions with the same phases (quantitative differences only) are labeled black.}
\label{difference}
\end{figure}

The quantitative and qualitative differences that are due to different interaction ranges are displayed visually in Fig.~\ref{difference}. Domains that are labeled white indicate parameter regions where different interaction ranges result in different strategy rankings (qualitatively different phases), while domains that are labeled black indicate regions with solely quantitative differences between the otherwise identical phases. To explain these surprising differences, it is instructive to start with the well-mixed interactions. In this case, the invasions between the players depend only on their strategies. If we introduce spatial structure with ``pair'' interactions, then the qualitative differences are expected because each player has a limited range, and thus the invasions depends not only on the strategies, but also on the spatial configuration of players. These differences are thoroughly documented and expected, and they also exist in other evolutionary games, like for example in social dilemmas, where spatial reciprocity may help cooperators survive where under well-mixed conditions they would perish \cite{nowak_n92b}. All the differences depicted in Fig.~\ref{difference}, however, do not have the same origin, and the arguments that apply when going from well-mixed to structured populations no longer apply when going from pairwise to group interactions on structured populations. Results presented in Fig.~\ref{difference} evidence clearly that both quantitative and qualitatively differences are common when going from ``pair'' to von~Neumann interactions (top), as well as when going from von~Neumann to Moore interactions (bottom).

Instead of the known ``well-mixed versus spatial'' arguments, it is important to acknowledge the fact that increasing the interaction range leads to indirect multi-point interactions between players that may not be directly connected via the interaction network, and which in the absence of group interactions would not be involved in the same elementary invasion processes. The most important consequence of this fact is that the relation between two neighboring competing strategies will depend not only on the invasion rate between them as defined with the basic food web, but also on the presence (or absence) of third parties. Other direct neighbors, as well as next-nearest neighbors (depending on the interaction range), may critically affect and indeed modify the relations defined by the basic food web depicted in Fig.~\ref{foodweb}. To demonstrate the validity of this argument, we study the movement of invasion fronts between homogeneous domains of strategies, which are the main driving force behind spatial pattern formation.

\begin{figure}
\centerline{\epsfig{file=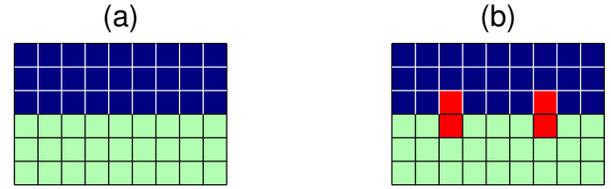,width=8cm}}
\caption{(Color online) Schematic presentation of two prepared initial states that we employ to demonstrate the importance of indirect multi-point interactions that arise due to group interactions. We focus on the movement of the invasion front between homogeneous domains of strategies, which is the main driving force behind spatial pattern formation. In panel (a), only two competing strategies, green (light gray) and blue (dark gray) are present, and hence the effect of multi-point interactions is less significant. 
This initial state is referred with subscript "2" in subsequent figures.
In panel (b), on the other hand, the interface is laced with players adopting the third strategy (red, medium gray) -- the third party -- that do not partake in strategy invasion, but they do interact with the other two competing strategies. 
We use subscript "3" in subsequent figures when the latter composition is used.
Most importantly, if group interactions govern the evolutionary process, the third party players will affect the payoffs and thus also the invasions between the green (light gray) and blue (dark gray) competing strategies. To maximize the latter effect when using Moore interactions, the fixed block of third party players is of the size $1 \times 4$, instead of the depicted $1 \times 2$ that we use for von~Neumann interactions. Colors used correspond to the $2 \to 1$ strategy invasion, but the same schematic presentation can of course be used for other examples as well. }
\label{prepared}
\end{figure}

More precisely, we consider two different prepared initial states, as illustrated in Fig.~\ref{prepared}. Firstly, shown in panel (a), only two competing strategies are present. Here we monitor the evolution of the frequency change of the dominant strategy -- the so-called excess frequency -- when starting from a symmetric initial state. In terms of the interaction range, we distinguish three different cases. In the first case, which we denote by $P_2$, only the two individuals involved in the invasion process interact with each other. In the second case players collect their payoff from the interactions with their four nearest neighbors (von~Neumann neighborhood, hence denoted by $N_2$). In the third case, we also consider the accumulation of payoff via the Moore neighborhood, which we denote by $M_2$. Secondly, shown in panel (b), we insert blocks of third-strategy players (red) along the interface. Importantly, the third party is not involved directly in the movement of the interface. The red players are fixed and cannot change their strategy, nor can they invade other players. Their only role is to influence the payoffs of the competing green and blue players if the conditions for this are given, i.e., in case of von~Neumann or Moore interactions. Similarly to the previously described initial state, here too we distinguish the same three interactions ranges, which we denoted by $P_3$, $N_3$ and $M_3$, respectively. Our goal is to determine how the presence of third-party players influences the invasion between the two competing predator-prey strategies.

\begin{figure}
\centerline{\epsfig{file=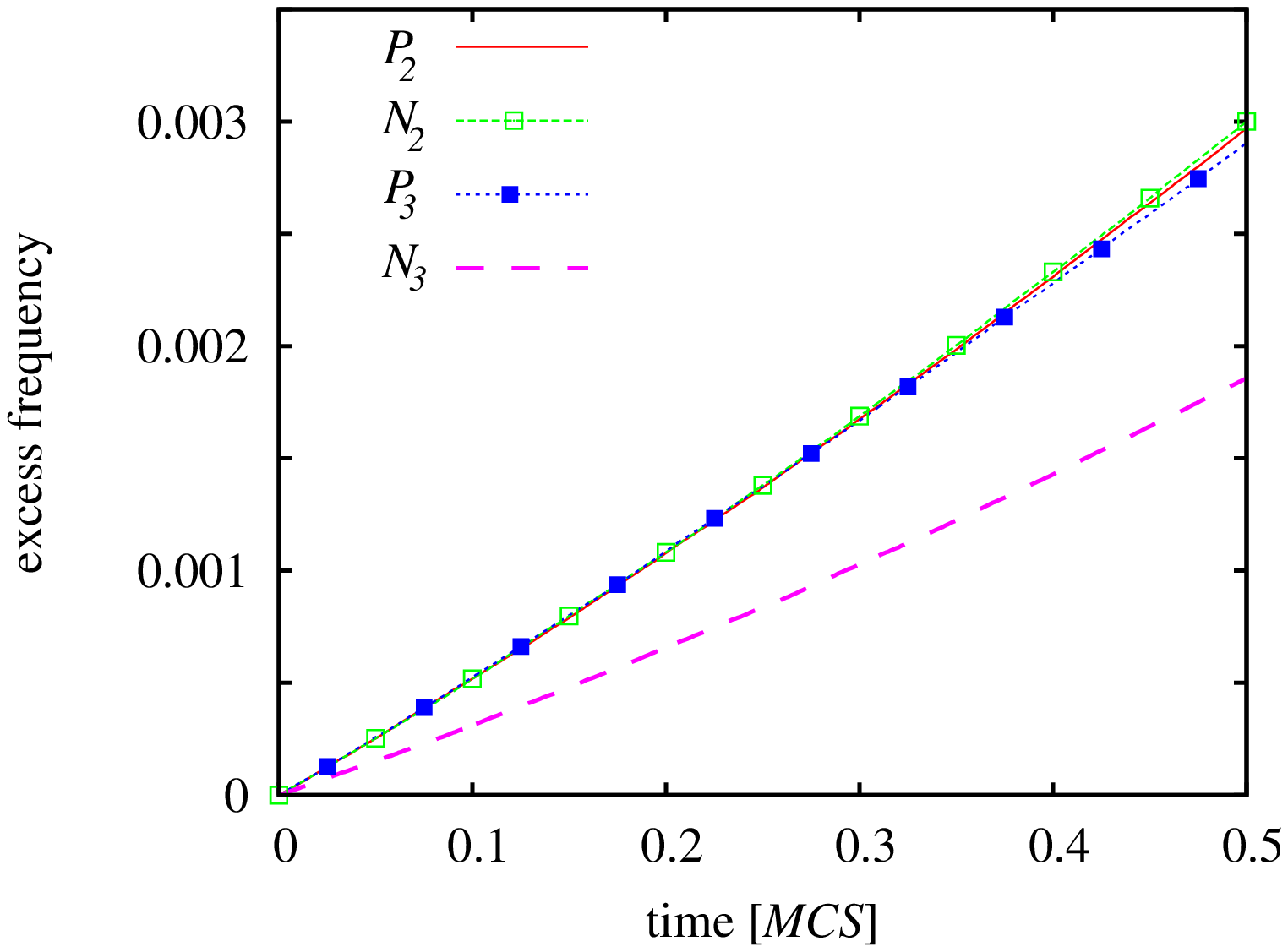,width=8.4cm}}
\centerline{\epsfig{file=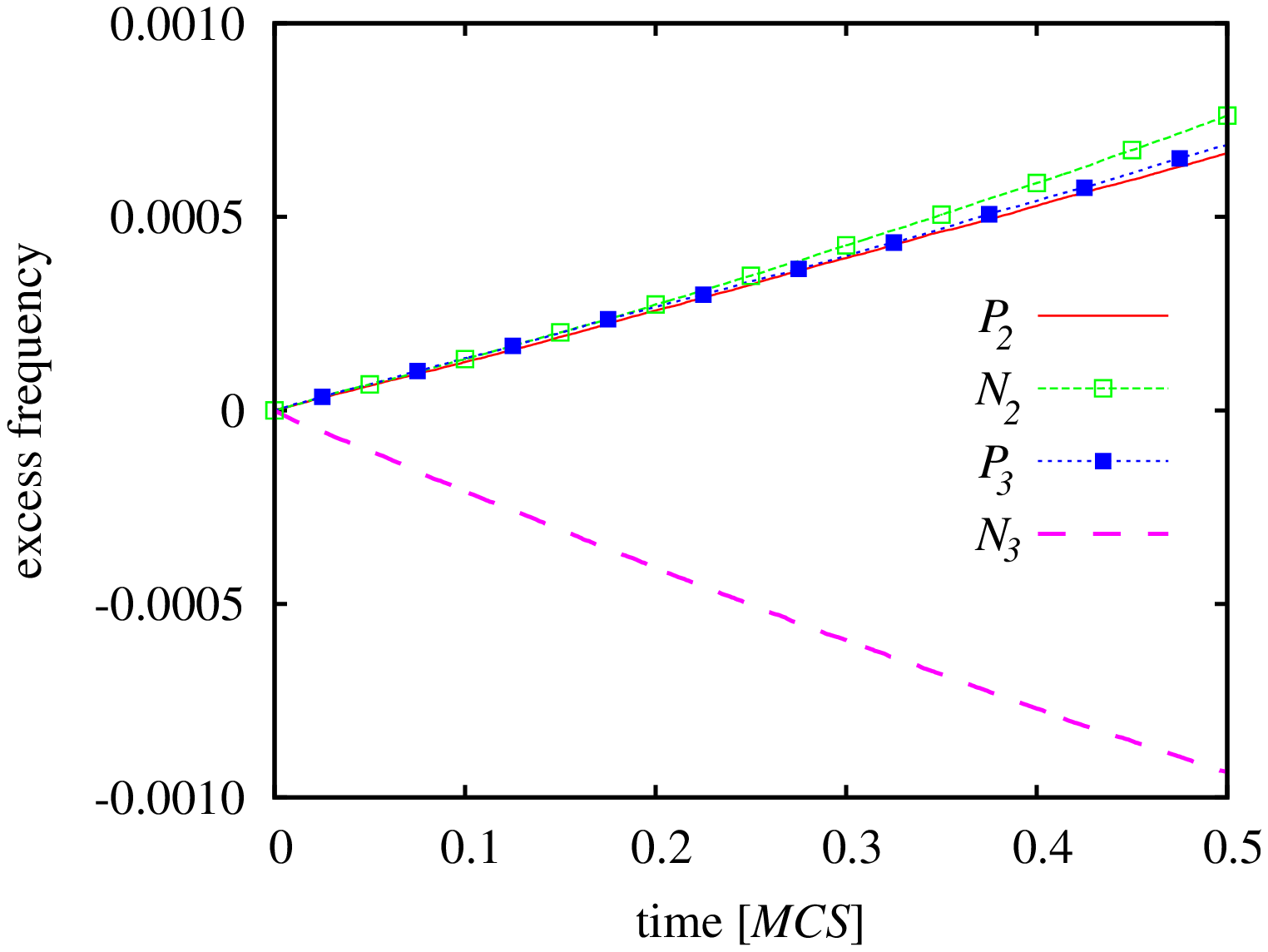,width=8.4cm}}
\caption{(Color online) Evolution of the excess frequency between two competing strategies, as obtained for $\delta_1=0.98$ and $\delta_0=0.25$. Upper panel shows the competition between the predator strategy $1$ and prey strategy $0$, while the lower panel shows the competition between the predator strategy $0$ and prey strategy $2$. In both panels, results obtained with the ``pair'' ($P_2$, $P_3$) and the von~Neumann ($N_2$, $N_3$) interactions are compared. It can be observed that quantitative (top panel) and qualitative (bottom panel) differences emerge as a consequence of the larger interaction range when third-party players are present ($N_2$ versus $N_3$). In case of ``pair'' interactions the differences are absent between the evolutions of $P_2$ and $P_3$. We note that the curves are scaled with the appropriate factor when the effective length of the invasion front is shortened by the fixed third-type players.}
\label{PN}
\end{figure}

In Fig.~\ref{PN}, we first compare the invasions between two predator-prey strategies when pair and von~Neumann interaction ranges are considered. As expected, if pair interactions govern the evolutionary process, there are no consequences stemming from the presence of the third strategy. Accordingly, the $P_2$ and $P_3$ curves are identical in both panels. The early evolution in case of von~Neumann interactions without third-party players (the $N_2$ curve) is also almost the same as the $P_2$ and $P_3$ case, because sideward and backward neighbors give zero contribution to the payoffs of the players who are at the front of the invasion. The critical impact of the third-party players' presence becomes obvious, however, when von~Neumann interaction range is applied. Here the multi-point interactions that are due to the extended interaction range take full effect. In the upper panel, the third strategy lowers significantly the effective invasion velocity, thus leading to a slower rise of the excess frequency of the predating strategy. It is important to note that this effect alone is capable to shuffle the rank of strategies, and thus qualitatively affect the phases, because changing a single interaction rate could influence the whole system due to cyclic dominance. But even more strikingly, as depicted in the lower panel, multi-point interactions alone are able to reverse the direction of invasion. Predators thus become prey solely because of the nearby presence of a third strategy. We note that group interactions do not always have such a dramatic effect on the outcome of an evolutionary process, and for some $\delta_1 - \delta_0$ pairs the effective invasion velocity between strategy domains remains practically unchanged, and the rank between species is preserved. The most striking effects can be observed in the regions labeled white in Fig.~\ref{difference}.

\begin{figure}
\centerline{\epsfig{file=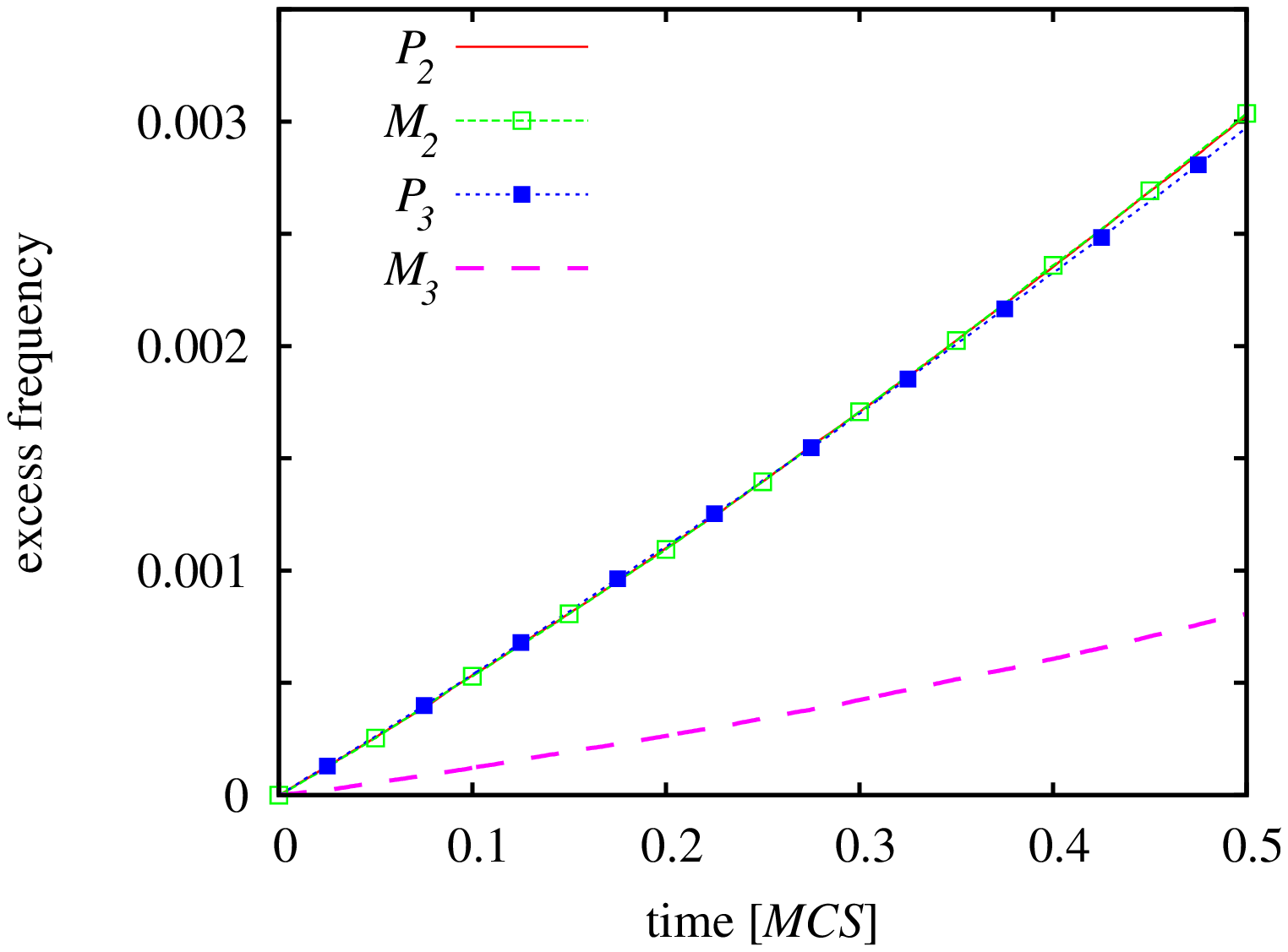,width=8.4cm}}
\centerline{\epsfig{file=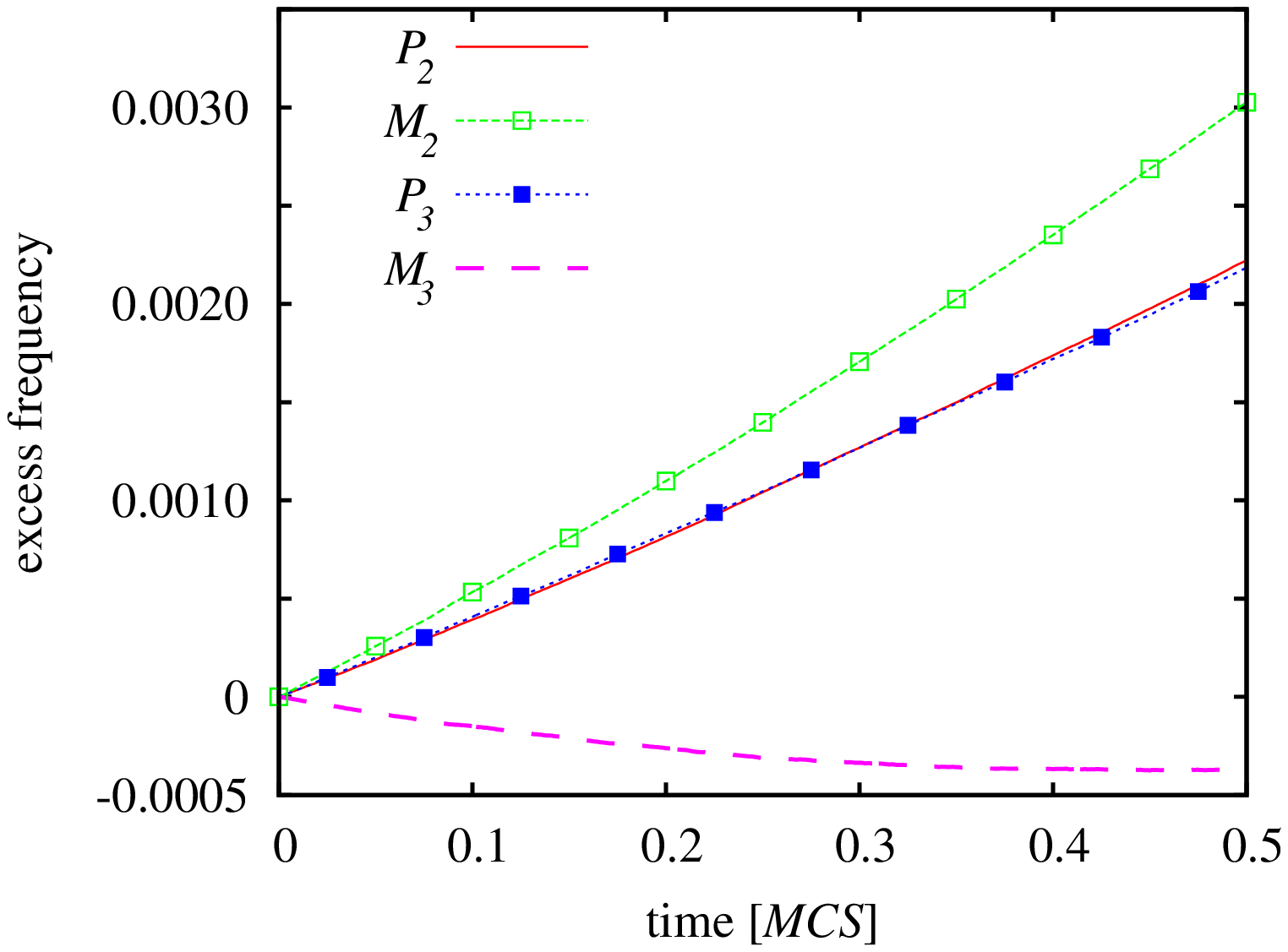,width=8.4cm}}
\caption{(Color online) Evolution of the excess frequency between two competing strategies, as obtained for $\delta_1=0.70$ and $\delta_0=0.76$. Up panel shows the competition between the predator strategy $2$ and prey strategy $1$, while the lower panel shows the competition between the predator strategy $0$ and prey strategy $2$. In both panels, results obtained with the ``pair'' ($P_2$, $P_3$) and the Moore ($M_2$, $M_3$) interactions are compared. As in Fig.~\ref{PN}, it can be observed that quantitative (top panel) and qualitative (bottom panel) differences emerge as a consequence of the larger interaction range when third-party players are present ($M_2$ versus $M_3$). In case of ``pair'' interactions $P_2$ and $P_3$ evolve similarly. We note that the curves are scaled with the appropriate factor when the effective length of the invasion front is shortened by the fixed third-type players.}
\label{PM}
\end{figure}

The same effects can also be observed if we compare the outcome of pair and Moore interactions, as depicted in Fig.~\ref{PM}. Results presented in the upper panel illustrate that the deceleration of the invasion front of predators can be even more spectacular. In the lower panel, the reversal of the direction of invasion is also clearly inferable. Furthermore, the even larger interaction range that is defined by the Moore neighborhood reveals another effect of group interactions. Namely, in the lower panel, there is obvious difference between the $M_2$ and $P_2$ (and $P_3$) curve. The early difference is due to the fact that players at the front of the invasion are, unlike in case of the von~Neumann interaction, now able to pick up extra payoff from more players of the front line due to the extended interaction range. This in turn accelerates the invasion of the predators and thus shifts the $M_2$ curve above the $P_2$ and the $P_3$ reference case.

Importantly, strategy configurations (not shown here) demonstrate that the same invasion rates between individual strategies can manifest in significantly different effective invasion rates between competing domains, which ultimately results in easily observable differences between the emerging spatial patterns. Finally, we note that we have also tested the robustness of our observations by using different host lattices. For example, we have used the hexagonal lattice, where every player has six direct neighbors and which is frequently used in biologically motivated settings \cite{hwang_cmb11}, and we have found that this lattice yields phase diagrams that are very similar to those that we have obtained for the square lattice in Fig.~\ref{phase}.

\section{Discussion}
We have studied the rock-paper-scissors game on a square lattice, thereby focusing on the transition from pairwise to group interactions governing the evolution of strategies. In addition to the expected effects of spatiality, which may introduce quantitative and qualitative changes in the evolutionary outcomes if compared to the well-mixed case, we have shown that the extent of the interaction range can have likewise profound consequences. We have demonstrated that increasing the interaction range from individual pairwise interactions to von~Neumann interactions and further to Moore interactions can both decelerate the propagation of predator strategy as well as revert the direction of invasion contrary to its definition by the governing food web. These results cannot be attributed to the traditional arguments that describe the transition from well-mixed to structured populations. Instead, we have shown that the key to understanding the emergent phenomena when going from pairwise to group interactions lies in the indirect multi-point interactions between players that are due to the extended interaction range. These effective multi-point interactions link together players that do not meet directly, and which in the absence of group interactions would not be involved in the same elementary invasion processes. We have validated these arguments by studying the movements of invasion fronts between homogeneous domains of strategies, which are arguably the main driving force behind spatial pattern formation. Based on the monitoring of the excess frequency of predators, we have confirmed both the deceleration of invasion fronts as well as the reversal of the direction of invasion. In conclusion, we have shown that group interactions can have a profound impact on the outcome of cyclic dominance games, even exceeding the impact reported before for evolutionary social dilemmas, where in the absence of strategic complexity solely the impact of noise becomes independent of the topology of the interaction network \cite{szolnoki_pre09c}.

Patterns resulting from cyclic dominance are common in nature, ranging from bacteria \cite{durrett_jtb97, kirkup_n04, nahum_pnas11} to plants \cite{taylor_am90, silvertown_je92} to the scale of ecological systems \cite{sinervo_n96, burrows_mep98}. Several previous works have illustrated that invasion rates between competing species can influence the resulting morphology \cite{law_je97}. Conversely, studying the morphology can be useful to deduce the microscopic invasion rates between species if they are not known, for example when bacteria or plants struggle for space. Our model illustrates, however, that conclusions stemming from such a procedure can be misleading, since the invasion rates alone do not determine the final fate of the competing species and their morphology. Moreover, in biological systems, it is widely accepted to measure the invasion rates or competition coefficients directly between species \cite{thorhallsdottir_je90, law_je97}. The presence of a third party, however, can result not only in a quantitative change of the invasion velocity, but may also altogether change the direction of invasion. Naturally, the weight of these arguments increases further if the number of competing species increases, like for example in bacterial warfare where more than one toxin is present \cite{szabo_jtb07}. Based on the presented results, we emphasize that it is important not to overlook the interaction range, which could also be a key parameter that renders reverse engineering cyclical interactions a very difficult undertaking.

\begin{acknowledgments}
This research was supported by the Hungarian National Research Fund (Grant K-101490), TAMOP-4.2.2.A-11/1/KONV-2012-0051, John Templeton Foundation (FQEB Grant \#RFP-12-22), and the Slovenian Research Agency (Grant J1-4055).
\end{acknowledgments}

\end{document}